\documentstyle[aps,graphicx,floats]{revtex}

\begin{document}

\draft
\wideabs{

\title{
High real-space resolution measurement of the local structure of
Ga$\bf _{1-x}$In$\bf _{x}$As using x-ray diffraction}

\author{ 
V. Petkov$^1$, I-K. Jeong$^1$, 
J. S. Chung$^1$, M. F. Thorpe$^1$, S. Kycia$^2$ and S. J. L. Billinge$^1$
}
\address{
$^1$Department of Physics and Astronomy and Center for
Fundamental Materials Research,
Michigan State University, East Lansing, MI 48824-1116.
$^2$Cornell High Energy Synchrotron Source,
Cornell University,
Ithaca, NY 14853}

\date{\today}

\maketitle

\begin{abstract}
High real-space resolution atomic pair  distribution 
functions (PDF)s from the alloy series Ga$_{1-x}$In$_x$As have been obtained using
high-energy x-ray diffraction. The first peak in the PDF is resolved as a doublet due to 
the presence of two nearest neighbor bond lengths,
Ga-As and In-As, as previously observed using XAFS.  The widths of nearest, and higher, neighbor pairs
are analyzed by separating the strain broadening from 
the thermal motion. The
strain broadening is five times larger for distant atomic neighbors as
compared to nearest neighbors. The results are in agreement with model 
calculations.
  
\end{abstract}
}

%\pacs{get new pacs}
%\pacs{71.30+h,71.38.+i,61.12-q,72.80.Ga}

%c% Semiconductor alloys, 
The average atomic arrangement of crystalline semiconductor alloys is
usually obtained from the position and intensities of the Bragg peaks in a
diffraction experiment~\cite{wycko;bk67}, 
and the actual nearest neighbor and sometimes next nearest
neighbor distances for various pairs of atoms by XAFS measurements~\cite{mikke;prl82}. 
In this Letter we show how high energy x-ray diffraction
and the resulting high-resolution atomic pair distribution functions (PDF)s
can be used for studying the internal strain in Ga$%
_{1-x}$In$_{x}$As alloys. We show that the first peak in the PDFs can be
resolved as a doublet and, hence, the mean position and also the widths of the Ga-As and
In-As bond length distributions determined. The detailed structure in the
PDF can be followed out to very large distances and the widths of the
various peaks obtained. We use the concentration dependence of the peak
widths to separate the strain broadening from the thermal broadening. At
large distances the strain broadening is shown to be about five times
larger than for nearest neighbor pairs. Using a simple valence
force field model, we get good agreement with the experimental results.

Ternary semiconductor alloys, in particular Ga$_{1-x}$In$_{x}$As, have
technological significance because they allow important properties, such as
band-gaps, to be tuned continuously between the two end-points by varying
the composition $x$. Surprisingly, there is no complete experimental
determination of the microscopically strained structure of these alloys.
On average, both\
GaAs and InAs form in the zinc-blende structure
 where Ga or In and As atoms occupy two inter-penetrating
face-centered-cubic lattices and are tetrahedrally coordinated to each
other~\cite{wycko;bk67}. However, both extended x-ray
absorption fine structure (XAFS) experiments~\cite{mikke;prl82} and
theory~\cite{cai;prb92ii} have shown
that Ga-As and In-As bonds do not take some average value but remain close to
 their 
{\it {natural}} lengths of $L^o_{\rm Ga-As}=2.437$~\AA\ and $L^o_{{\rm In-As}}=
2.610$~\AA\ in the alloy. 
Due to the two considerably different bond lengths
present, the zinc-blende structure of Ga$_{1-x}$In$_{x}$As\ alloys becomes
locally distorted. A number of 
authors~\cite{mikke;prl82,cai;prb92ii,marti;prb84,balza;prb85} 
have proposed distorted local structures
but there has been limited experimental data available to date.  The fully
distorted structure is a prerequisite as an input for accurate band structure
and phonon dispersion calculations~\cite{zunge;prl90}.

The technique of choice for studying the local structure of semiconductor alloys has been
XAFS~\cite{mikke;prl82,balza;prb85}.
However, XAFS provides information only about the immediate atomic ordering
(first and sometimes second coordination shells) and all longer-ranged
structural features remain hidden. To remedy this shortcoming we have taken
the alternative experimental approach of obtaining high-resolution PDFs of
these alloys from high energy x-ray diffraction data.

\begin{figure}[!tb]
  \centering
  \includegraphics[angle=0,width=2.8in]{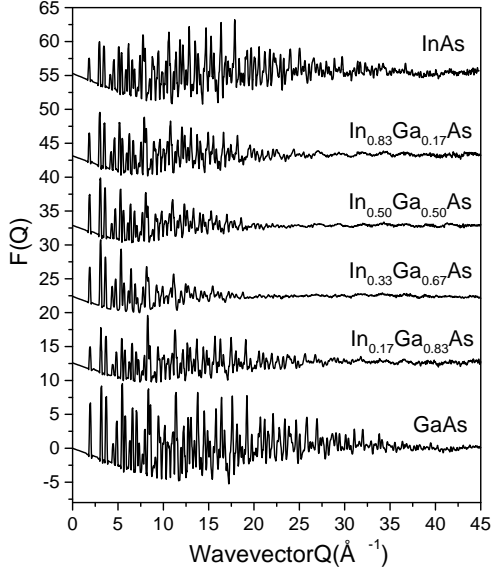}
%c%  \caption{Reduced structure factor $F(Q)=Q[S(Q)-1]$ for 
  \caption{The reduced structure factor $F(Q)=Q[S(Q)-1]$ for Ga$_{1-x}$In$_{x}$%
    As measured at 10K for various concentrations.}
  \label{fig;sq}
\end{figure}

\begin{figure}[!tb]
  \centering
  \includegraphics[angle=0,width=2.8in]{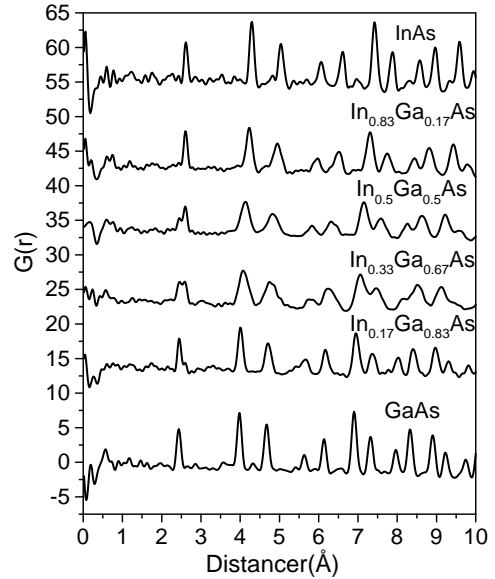}
%c%  \caption{Reduced radial distribution function, G(r), for 
  \caption{The reduced PDF, $G(r)$, for Ga$_{1-x}$In$_{x}$As measured at 10~K
   for various concentrations.}
  \label{fig;gr}
\end{figure}

The PDF is the instantaneous atomic density-density correlation function
which describes the local arrangement of atoms in a material~\cite
{warre;bk90}. It is the sine Fourier transform of the experimentally
observable total structure function obtained from powder diffraction
measurements. 
PDF analysis yields the real local structure whereas an analysis of the
Bragg scattering alone yields the average crystal structure. Determining the PDF
has been the approach of choice for characterizing glasses, liquids and
amorphous materials for a long time~\cite{wased;bk80}. However,
its widespread application to study {\it {crystalline}} materials has been
relatively recent~\cite{egami;mt90}. 
Very high real-space resolution is
required to differentiate the distinct Ga-As and In-As bond lengths present
in Ga$_{1-x}$In$_{x}$As. 
High real-space resolution is obtained by measuring the
structure function, $S(Q)$ ($Q$ is the amplitude of the wave vector), to a
very high value of $Q$ ($Q_{\max }\geq 40$~\AA $^{-1}$).  An indium
neutron absorption resonance rules out neutron measurements in the 
Ga$_{1-x}$In$_{x}$As system. We therefore carried out x-ray powder
diffraction measurements.  To access $Q$
values in the vicinity of 40-50~\AA $^{-1}$ it is necessary to use x-rays
with energies $\geq 50$~keV. 
The experiments were carried out at the A2 56 pole wiggler beamline at
Cornell High Energy Synchrotron Source (CHESS)
which is capable of delivering intense x-rays of energy 60 keV. Six powder
samples of Ga$_{1-x}$In$_{x}$As, with $x=0.0$, 0.17, 0.5, 0.67, 0.83 and
1.0, were measured. The samples were made by standard methods and the
details of the sample preparation will be reported elsewhere~\cite
{petko;prb99;unpub}. All measurements were done in symmetric transmission
geometry at 10K. Low temperature was used to minimize thermal vibration
in the samples, and hence to increase the sensitivity to atomic static
displacement amplitudes. A double crystal Si(111) monochromator was used. 
Scattered radiation was collected with an intrinsic
germanium detector connected to a multi-channel analyzer.
The elastic component was separated from the incoherent Compton scattering
before data analysis~\cite{petko;prb99;unpub}. 
Several diffraction runs were conducted with each sample
and the resulting spectra averaged to improve the statistical accuracy.
The data were normalized for flux, corrected for background scattering,
detector deadtime and absorption and divided by the average form factor to
obtain the total structure factor, 
$S(Q)$~\cite{warre;bk90,wased;bk80,klug;bk74}. 
Details of the data processing are
described elsewhere~\cite{petko;prb99;unpub}. 
Correction procedures were done using the program 
RAD~\cite{petko;jac89}. Experimental reduced structure factors, 
$F(Q)=Q[S(Q)-1]$, are
shown in Fig.~\ref{fig;sq}. 
%
%c% The corresponding reduced radial distribution functions , $G(r)$, 
The corresponding reduced PDFs, $G(r)$, obtained through a Fourier
transform %c% of $\ S(Q)$ according to $G(r)=2/\pi

\begin{equation}
G(r)={\frac{2}{\pi }}\,\int_{0}^{Q_{max}}F(Q)\sin Qr\,dQ
\end{equation}
are shown in Fig.~\ref{fig;gr}. 
The data for the Fourier transform were terminated at $Q_{max}=45$~\AA $^{-1}
$ beyond which the signal to noise ratio became unfavorable. This is a very
high momentum transfer for x-ray diffraction measurements; for comparison,
$Q_{max}$ from a Cu K$_{\alpha }$ x-ray tube which is less than 8~%
\AA $^{-1}$.

\begin{figure}[!tb]
  \centering
  \includegraphics[angle=0,width=2.8in]{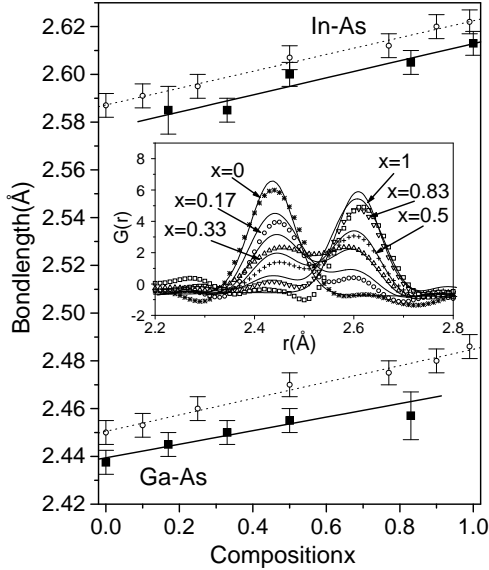}
 \caption{Solid symbols: Ga-As and In-As bond-lengths vs.
 composition as extracted from the present PDFs. Open symbols:
room-temperature XAFS results from Ref.~\protect\cite{mikke;prl82}. 
Inset: Split nearest neighbor PDF peak from the data (symbols) and the model
(solid lines).}
\label{fig;zplot}
\end{figure}

Significant Bragg scattering (well-defined peaks) are immediately evident in
Fig.~1 up to $Q\sim 40$~\AA $^{-1}$\ in the end-members, GaAs and InAs. This
implies that the samples have long range order and that there is little
positional disorder (dynamic or static) on the atomic scale. The Bragg-peaks
disappear at much lower $Q$-values in the alloy data: the samples are
still long-range ordered but they have significant local positional disorder. 
At high-$Q$ values, oscillating diffuse scattering is
evident. This has a period of $~2\pi /2.5$~\AA $^{-1}$ and contains
information about the shortest atomic distances in Ga$_{1-x}$In$_{x}$As
alloys seen as a sharp first peak in $G(r)$ at 2.5~\AA\ (see Fig.~\ref{fig;gr}). 
In the
alloys, this peak is split into a doublet as is clearly evident in Fig.~\ref{fig;gr};
with a shorter Ga-As bond and a longer In-As bond. This peak is shown on an
expanded scale in the inset to Fig.~\ref{fig;zplot} for all the compositions
measured.  We determined the positions of the two subcomponents of the first
PDF peak, i.e. the mean Ga-As and In-As bond lengths, and the results are shown
in Fig.~\ref{fig;zplot}. Also shown is the room temperature result
previously obtained in the XAFS study of Mikkelson and Boyce~\cite
{mikke;prl82}. There is clearly good agreement. The PDF-based bond lengths
are shifted to smaller lengths by about 0.012~\AA\ since our data were
measured at 10K, whereas the XAFS experiments were at room temperature. 
The nearest neighbor peak is the only peak which is sharp in the
experimental PDFs as can be seen in Fig.~2. From the second-neighbor onwards
the significant strain in the alloy samples results in broad atom-pair
distributions without any resolvable splitting. Model calculations show
that this broadening is intrinsic and not due to any experimental
limitations.
The strain in Ga$_{1-x}$In$_{x}$As was quantified by fitting the
individual peaks in experimental PDFs. We used Gaussians convoluted with 
Sinc functions which account for the experimental resolution coming from
the finite $Q_{max}$. The FWHM of the resolution function is
0.086~\AA .  This is significant for the near-neighbor peaks as shown
in Fig.~3, but is much smaller than the width of the high-$r$ peaks.
The high-r peaks are fit using the PDFFIT modeling
program\cite{proff;jac99} assuming the virtual crystal zinc-blende structure
and refining displacement parameters.
The resulting mean-square Gaussian standard deviations
are shown in Fig. 4. One can see Ga-As and 
In-As bond lengths are sharply peaked 
about the mean value whereas the static strain on more distant neighbors 
is five times larger than on the near-neighbors. The 
strain peaks at a composition $x$=0.5 and
affects the common (As) more than the mixed (metal) sublattice. 

In order to better understand these results, we have modeled to the static
and thermal disorder in the alloy by using a Kirkwood potential~\cite{kirkwood}. The
key element in this potential is the central force term that connects
nearest neighbor atoms and would like to keep each bond at its natural ({\it {%
unstrained}}) length. The potential contains nearest neighbor bond
stretching force constants $\alpha $ and force constants $\beta $ that
couple to the change in the angle between adjacent nearest neighbor bonds.
We choose these parameters to fit the end members \cite{cai;prb92ii} with $\alpha _{{\rm {%
Ga-As}}}$ = 96N/m, $\alpha _{{\rm {In-As}}}$ = 97N/m, $\beta _{{\rm {Ga-As-Ga%
}}}$ = $\beta _{{\rm {As-Ga-As}}}$ = 10N/m and $\beta _{{\rm {In-As-In}}}$ = 
$\beta _{{\rm {As-In-As}}}$ = 6N/m. The additional angular force constant
required in the alloy are taken to be the geometrical mean, so that $\beta _{%
{\rm {Ga-As-In}}}$ = $\sqrt{(\beta _{{\rm {Ga-As-Ga}}}.\beta _{{\rm {In-As-In%
}}})}$. We have constructed a series of cubic 512 atom periodic supercells
in which the Ga and In atoms are distributed randomly according to the
composition $x$. The system is relaxed using the Kirkwood potential to find
the displacements from the virtual crystal positions. The volume of the
supercell is also adjusted to find the minimum energy. Using this strained
static structure, a dynamical matrix has been constructed and the
eigenvalues and eigenvectors found numerically. From this the Debye-Waller
factors for all the individual atoms in the supercell can be found and hence
the PDF of the model
by including the Gaussian broadening of all the subpeaks. We have
shown previously~\cite{TC} that this is the correct procedure within the harmonic
approximation.  The model-PDF is plotted with the data
in the inset to Fig.~2 and in Fig.~5. The agreement at higher-$r$ is
comparable to that in the $r$-range shown.
All the peaks shown in the Figures
consist of many Gaussian subpeaks. 
The overall fit to the experimental
$G(r)$ is excellent and the small discrepancies in Fig.~5
between theory and experiment are probably due to small residual
experimental errors. Note that in comparing with experiment, the theoretical PDF has 
been convoluted with a sinc function to incorporate the truncation of the experimental data 
at $Q_{max} = 45$~\AA . The technique discussed above could be extended using a better
force constant model with more parameters, but does not seem necessary at
this time.

The thermal and strain contributions to the widths of the individual peaks
in the reduced PDF act independently, as expected and as confirmed by our supercell
calculations described in the previous paragraph. We therefore expect the squared width $%
\Delta $ to be a sum of the two parts. The thermal part $\sigma $ is almost
independent of the concentration and we fit $\sigma ^{2}$ by a linear
function of the composition $x$ between the two end points in Fig.~4. To
better understand the strain model it is convenient to assume that all the
force constants are the same and independent of chemical species. Then it
can be shown \cite{Rosa} for any such model that

\begin{equation}
\Delta _{ij}^{2}=\sigma _{ij}^{2}+A_{ij}x(1-x)(L^o_{\rm In-As}-L^o_{\rm Ga-As})^{2}
\end{equation}
where the subscripts $ij$ refer to the two atoms that lead to a given peak
in the reduced PDF. For the Kirkwood model the $A_{ij}$ are functions of the
ratio of force constants $\beta /\alpha $ only. It further turns out that
the $A_{ij}$ are independent of whether a site in one sublattice is Ga or
In, so we will just refer to that as the metal site. Taking mean values from
the force constants used in the simulation we find that $\beta /\alpha $ =
0.83, and that for nearest neighbor pairs $A_{ij}$ = 0.0712. For more
distant pairs the motion of the two atoms becomes incoherent so that $%
A_{ij}=A_{i}+A_{j}$ and we find that for the metal site $A_{i}=0.375$ and
for the As site $A_{i}=1.134$. The validity of the approximation of
using mean values for the force constants was shown to be accurate
 by calculating the model-PDF for all compositions as described 
above and comparing to the prediction of Eq. (2)~\cite{Rosa}.
Equation~2 shows good agreement with the
data for near and far neighbor PDF peaks, and for the different sublattices,
over the whole alloy range, as shown in Fig.~4, using only parameters taken from fits to 
the end-members. There is a considerably larger width associated with the As-As peak in 
Fig.~4 when compared to the Me-Me peak, because the As atom is surrounded by four
 metal cations, providing five distinct first-neighbor environments \cite{marti;prb84,balza;prb85}. The theoretical 
curve in the lower panel of Fig.~4 is predicted to be the same for the Ga-As and In-As bond length distribution, using the simplified
approach.  The Kirkwood model seems adequate to describe the experimental data at this time, although further refinement of the error 
bars may require the use of a better potential containing more parameters.

\begin{figure}[!tb]
  \centering
  \includegraphics[angle=0,width=2.8in]{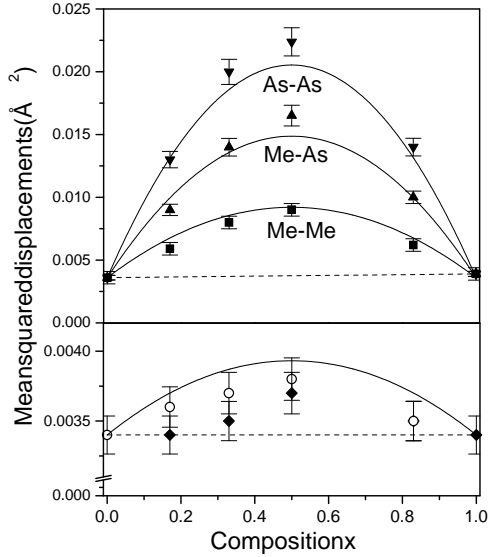}
\caption{square of the PDF peak widths for far neighbors (top panel) and
nearest neighbors (lower panel) separated by sub-lattice type. 
Symbols: values from the data. In the lower panel the open symbols are
for the Ga-As and the closed symbols the In-As bond. See text for details. 
Solid lines: theory.  The mean-square static and thermal distortions are added.
Here Me represents both the metals Ga and In, which
behave in the same way. Note the scale in the lower panel is expanded by a factor 10 compared to the upper panel.}
\label{fig;parabolas}
\end{figure}
\begin{figure}[!tb]
  \centering
  \includegraphics[angle=0,width=2.8in]{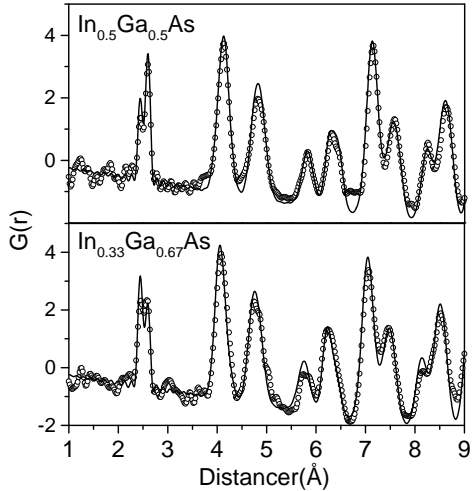}
  \caption{Experimental (open circles) and theoretical (solid line) PDFs for
   Ga$_{1-x}$In$_{x}$As for concentrations  $x=0.5$ and $x=033$.}
 \label{fig;fits}
\end{figure}

In summary, we report for the first time a high-real-space-resolution
measurement of the PDF of Ga$_{1-x}$In$_{x}$As ($0<x<1$) alloys. The PDF
allows the local distortions away from the average structure over a wide
range of $r$ to be
determined in disordered crystals such as these.  The
nearest-neighbor Ga-As and In-As bond lengths in the alloys are clearly resolved.
Significantly greater  disorder exists in the more distant neighbor length 
distributions in the
alloys. The experimental results are well fit over a wide range or $r$
using a Kirkwood model.  Because the agreement between theory and experiment 
is good at both short and large distances, the Kirkwood model can be used with some confidence to
generate strained alloy structures for use in the calculation of electronic band structure 
and phonon dispersion curves.

We would like to thank Rosa Barabash for discussions and help with the
analysis of the static strains and Andrea Perez and the support staff at
CHESS for help with data collection and analysis. This work was supported by
DOE through grant DE FG02 97ER45651. CHESS is operated by NSF through grant
DMR97-13424.

%\bibliography{/u24/billinge/bib/hts,%
%	      /u24/billinge/bib/htsii,%
%	      /u24/billinge/bib/ndb,%
%	      /u24/billinge/bib/1995,%
%	      /u24/billinge/bib/1996,%
%	      /u24/billinge/bib/1997,%
%	      /u24/billinge/bib/1998,%
%	      /u24/billinge/bib/emil,%
%	      /u24/billinge/bib/1999,%
%	      }
%\bibliographystyle{/u24/billinge/bib/aip_simon}
%\bibliographystyle{/u24/billinge/tex/bib/aip}

\end{document}